\documentclass{elsart}

\usepackage{graphicx}
\begin{document}
\begin{frontmatter}
\title{What is Surrealistic about Bohm Trajectories?}

\author{M.O. Terra Cunha}

\address{Instituto de Ci\^encias Exatas \\
	Universidade Federal de Minas Gerais \\
	CP 702, 30123-970, Belo Horizonte, MG, Brazil}

\thanks{e-mail: tcunha@mat.ufmg.br \\
	fax: 55 31 499-5797}

\begin{abstract}
 We discuss interferometers in Bohmian quantum mechanics.
 It is shown that, with the correct configuration space,
 Bohm trajectories in a which way interferometer are not
 \emph{surrealistic}, but behaves exactly as
 common sense suggests. Some remarks about a way to generalize
 Bohmian mechanics to treat density matrix are also made.
 
\end{abstract}


\end{frontmatter}

\section{Introduction}

Orthodox quantum mechanics, in the sense of Copenhagen interpretation, has
succeed very well in its role of taking account experimental results. It is
a kind of common sense that orthodox quantum mechanics is \emph{fapp}, i.e.
it is good \emph{for all practical purposes}.

But it is also almost common sense that in the scope of Copenhagen
interpretation it is not possible to \emph{understand} quantum mechanics\cite
{Fey65}. Just in the green years of quantum theory, one of his fathers, de
Broglie, proposed\cite{dBr56} the \emph{pilot wave} interpretation. One
great problem of this attempt was that de Broglie pilot waves live in
ordinary physical space, what made the generalization to many body systems a
little fuzzy.

Another attempt to insert trajectories in quantum mechanics was made by Bohm%
\cite{Bohm52}, in 1952. In his brilliant interpretation, the wave field
actually creates a \emph{quantum potential}, and particle trajectories are
obtained as rays of a Hamilton-Jabobi like equation, where the quantum
potential is added to the classical one. Two elements play decisive roles in
Bohmian mechanics: the wave field $\Psi $ and configuration space, where
Bohm trajectories live.

Up to now (and to author's knowledge), Bohmian mechanics has survived to all
critics, except perhaps for the one which will be treated in this letter.

In the year of 1992, when Bohmian mechanics completes 40 years and his
father passed away, Englert, Scully, S\"{u}ssmann, and Walther gave to the
community the striking article\cite{ESSW92} entitled \emph{Surrealistic Bohm
Trajectories}. In this work, authors considered one bit \emph{which way} 
(WW, also
for \textit{welcher Weg}) interferometers' thought experiments (but almost
realizable with modern quantum optics technology) and Bohm trajectories of
simple (in opposition to WW) interferometers\cite{PDH79} to construct an
incongruence between common thought and Bohm trajectories. With this in
hands, referred authors claimed Bohm trajectories could not be realistic,
and should be ``surrealistic''.

This example of ``surrealistic'' Bohm trajectories was criticized in many
ways, some of them claimed that between Bohmian mechanics and common sense,
the first is stronger\cite{DFGZ93}, other suggested there should be
mathematical mistakes\cite{Scu98}, and other even appealed to nonlocal
instantaneous teleportation of energy to save the appearances\cite{DHS93}.

The intention of this letter is to reanalyze ESSW experiment (in fact a
pedagogical modification suggested by Dewdney, Hardy, and Squires\cite{DHS93}%
) but taking account of the role played by configuration space in Bohmian
mechanics. In this way we show there is nothing ``surrealistic'' about Bohm
trajectories in the example worked, and even that they agree with common
sense when worked in the right way. Once again, it can be viewed as a feature
of Bohr's insistence on the essential wholeness of quantum mechanics.

This letter is organized as follows: we first review interferometers in
Bohmian mechanics and sketch ESSW argument, then we review WW
interferometers in the orthodox approach, and finally we revise the Bohmian
approach to these systems. We then close the letter with brief conclusions.

\section{Interferometers in Bohmian Mechanics and ESSW Argument}

In this section we discuss how to understand interferometers in Bohmian
mechanics. For simplicity we work with an incomplete (i.e. without the last
beam splitter) Mach-Zender interferometer (fig 1). As is the essence of any
interferometer, two ways are permitted and an interference ``region'' takes
place. Then, we sketch ESSW argument on ``surrealistic'' Bohm trajectories.

In Bohmian mechanics we write the wave field $\Psi \left( \mathbf{x}\right) $%
, where $\mathbf{x}$ denotes a configuration space parametrization, in polar
representation 
\[
\Psi =R\exp \left( \frac i\hbar S\right) , 
\]
and Schr\"{o}dinger equation implies\cite{Bohm52} $P=R^2$, and $S$ should
obey a continuity equation 
\[
\frac{\partial P}{\partial t}+\nabla \cdot \left( P\frac{\nabla S}m\right)
=0, 
\]
and a Hamilton-Jacobi like one 
\[
\frac{\partial S}{\partial t}+\frac{\left( \nabla S\right) ^2}{2m}+V\left( 
\mathbf{x}\right) +U\left( \mathbf{x}\right) =0, 
\]
where $U\left( \mathbf{x}\right) $ is the so called \emph{quantum potential}%
, given by 
\[
U\left( \mathbf{x}\right) =\frac{-\hbar ^2}{2m}\frac{\nabla ^2R}R. 
\]

After passage by the beam splitter, the wave field becomes (we shall neglect
normalization factors whenever possible) 
\[
\Psi \left( \mathbf{x}\right) =\psi _r\left( \mathbf{x}\right) +\psi
_t\left( \mathbf{x}\right) .
\]
Except by the region $I$, $\psi _r$ and $\psi _t$ do not overlap. The
quantum potential in this region is the same as for free particles (some
details about wave packet structure are also interesting, but are beyond the
scope of this letter). But for free particles $R$ is constant and quantum
potential exerts no effect! In region $I$, however, the interference pattern
is important and we actually have 
\[
R^2=R_r^2+R_t^2+2R_rR_t\cos \left( \frac{S_r-S_t}\hbar \right) \simeq
2R_r^2\left( 1+\cos \Phi \right) ,
\]
and quantum potential becomes very important because $\nabla ^2R$ grows up
and $R$ becomes small. That is an interesting point about Bohmian mechanics
which should be emphasized: \emph{interference} becomes \emph{interaction} 
through the
quantum potential!

\begin{figure}
\begin{picture}(400,200)(-60,-100)

%%%% It is possible that (5,4) be a better
%%%% choice for slope of rays!

\thinlines  % for rays
\put (-50,40){\line(5,-4){50}}% Reflected ray
\put (0,0){\line(5,4){70}}
\put (70,56){\line(5,-4){50}}
\put (160,-16){\line(5,-4){20}}% Reflected ray ended

\put (-50,40){\line(5,-4){120}}% Transmitted ray
\put (70,-56){\line(5,4){50}}
\put (160,16){\line(5,4){20}}% Transmitted ray ended

\put (140,0){\oval(24,40)}% I Region

\thicklines %for `optical' elements
\put (-10,0){\line(1,0){20}}% Beam Splitter
\put (60,57){\line(1,0){20}}% Upper Mirror
\put (60,-57){\line(1,0){20}}% Down Mirror
\put (191,28.5){\line(-4,5){12}}% Detector 1
\put (191,-28.5){\line(-4,-5){12}}% Detector 2

\put(20,28){\shortstack{$r$}}
\put(20,-40){\shortstack{$t$}}
\put(190,40){\shortstack{$1$}}
\put(190,-52){\shortstack{$2$}}
\put(136,-35){\shortstack{$I$}}

\end{picture}
\caption{Experimental scheme of an incomplete Mach-Zender interferometer.} %% FILE fig1.tex SHOULD BE AVAILABLE

\end{figure}
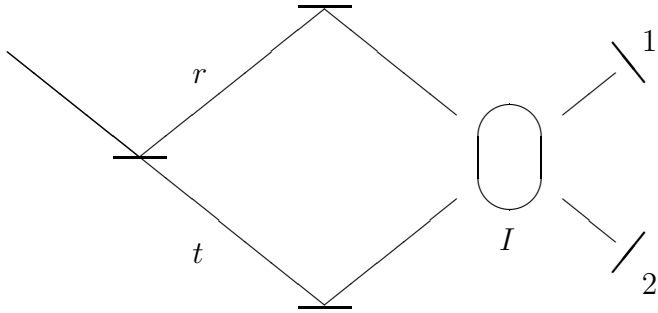

The role played by this interference region in Bohm trajectories is
dramatic. In \cite{PDH79} it can be found pictures of the quantum potential
and Bohm trajectories for double slit interferometer, which essence also
fills in the analysis made above.

In the simple example worked, if we forget details about Bohm trajectories
and analyze only the question: in which detector will the trajectory
which passes by $r$ $\left( t\right) $ end? we can understand the (at first time
strange) fact that this trajectory should end up in detector $1$ $\left(
2\right) $ (see fig. 2).

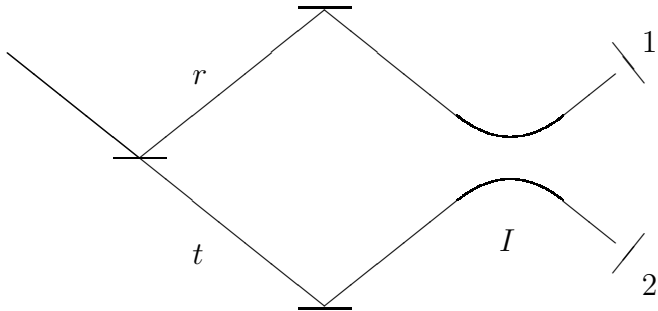
\begin{figure}
\begin{picture}(400,200)(-60,-100)

%\thicklines  % for trajectories
\put (-50,40){\line(5,-4){50}}% Reflected ray
\put (0,0){\line(5,4){70}}
\put (70,56){\line(5,-4){50}}
\put (160,-16){\line(5,-4){20}}% Reflected ray ended

\put (-50,40){\line(5,-4){120}}% Transmitted ray
\put (70,-56){\line(5,4){50}}
\put (160,16){\line(5,4){20}}% Transmitted ray ended

\qbezier(120,16)(140,0)(160,16)% r - 1
\qbezier(120,-16)(140,0)(160,-16)% t - 2
\thinlines %for `optical' elements
\put (-10,0){\line(1,0){20}}% Beam Splitter
\put (60,57){\line(1,0){20}}% Upper Mirror
\put (60,-57){\line(1,0){20}}% Down Mirror
\put (191,28.5){\line(-4,5){12}}% Detector 1
\put (191,-28.5){\line(-4,-5){12}}% Detector 2

\put(20,28){\shortstack{$r$}}
\put(20,-40){\shortstack{$t$}}
\put(190,40){\shortstack{$1$}}
\put(190,-52){\shortstack{$2$}}
\put(136,-35){\shortstack{$I$}}

\end{picture}
\caption{Bohm trajectories for an incomplete Mach-Zender interferometer.}  %% FILE fig2.tex SHOULD BE AVAILABLE

\end{figure}

Just by symmetry arguments (in a balanced interferometer) it follows $\psi
_r\left( \mathbf{x}\right) =\pm \psi _t\left( \mathbf{Rx}\right) $ where $%
\mathbf{R}$ denotes reflection of configuration space through the beam
splitter plane. Though $\Psi \left( \mathbf{Rx}\right) =\pm \Psi \left( 
\mathbf{x}\right) $, $R\left( \mathbf{Rx}\right) =R\left( \mathbf{x}\right) $%
, and $S\left( \mathbf{Rx}\right) =_{\mathrm{mod}h/2 }S\left( \mathbf{x}%
\right) .$ As in Bohmian mechanics the velocity field is given by 
\[
\mathbf{v=}\frac{\nabla S}m,
\]
we obtain that Bohm trajectories can not cross the plane of the beam
splitter in the \textit{free} region.

Just as another equivalent argument, Bohm trajectories are like flux lines
and so they can not cross one another. This establishes that Bohm
trajectories, \emph{for the simple interferometer}, are $r\rightarrow 1$, $%
t\rightarrow 2$.

ESSW argument can be viewed as follows: in orthodox quantum mechanics, there
is no interaction in region $I$, so particles just follow free evolution.
Free evolution takes the wave packet described by $\psi _r$ $\left( \psi
_t\right) $ to detector $2$ $\left( 1\right) $. So, if we can independently
record which way ($r$ or $t$) and which detector ($1$ or $2$) a one particle
wave packet marks, one can distinguish between Bohm and Copenhagen quantum
mechanics. Assuming the second to give the right result, Bohm trajectories
should be ``surrealistic''.

\section{WW Interferometers}

In this section we discuss the WW interferometers\cite{SEW91} in orthodox
quantum mechanical formalism.

We now change slightly our \textit{gedanken} experiment apparatus, following 
ESSW, by adding
a one bit which way detector. This can be realized, for example, if we
consider our beam as a beam of Rydberg circular atoms and in each of the
paths we include a microwave cavity resonant with just one mode decay of the
atom, and experimentally we made a velocity selection such that, whenever the
atom enters a cavity in the excited state it should emit a photon and leave
the cavity in the lower state (in quantum optics terminology, this is called
a $\pi $ pulse). This is shown in figure 3, as an example, but for our
purposes it is only necessary to include one bit information about which
way, independently of how it is realized (in the example this bit would be
in which cavity the photon was emitted). Mnemonically, this bit will assume
values $r$ or $t$.

\begin{figure}
\begin{picture}(400,200)(-60,-100)

%%%% It is possible that (5,4) be a better
%%%% choice for slope of rays!

\thinlines  % for rays
\put (-50,40){\line(5,-4){50}}%    Reflected ray
\put (0,0){\line(5,4){70}}
\put (70,56){\line(5,-4){50}}
\put (160,-16){\line(5,-4){20}}%   Reflected ray ended

\put (-50,40){\line(5,-4){120}}%   Transmitted ray
\put (70,-56){\line(5,4){50}}
\put (160,16){\line(5,4){20}}%     Transmitted ray ended

\put (140,0){\oval(24,40)}%        I Region

\thicklines %for `optical' elements
\put (-10,0){\line(1,0){20}}%       Beam Splitter
\put (60,57){\line(1,0){20}}%      Upper Mirror
\put (60,-57){\line(1,0){20}}%      Down Mirror
\put (191,28.5){\line(-4,5){12}}%  Detector 1
\put (191,-28.5){\line(-4,-5){12}}%Detector 2

\put(20,28){\shortstack{$r$}}
\put(20,-40){\shortstack{$t$}}
\put(190,40){\shortstack{$1$}}
\put(190,-52){\shortstack{$2$}}
\put(136,-35){\shortstack{$I$}}

% Now WW

\qbezier(34,39.5)(35.8,47.5)(44,47.5)% Cavity r
\qbezier(46,24.5)(54.2,24.5)(56,32.5)

\qbezier(34,-39.5)(35.8,-47.5)(44,-47.5)% Cavity t
\qbezier(46,-24.5)(54.2,-24.5)(56,-32.5)

\put(60,25){\shortstack{$\mathrm{C}_r$}}
\put(60,-37){\shortstack{$\mathrm{C}_t$}}
\end{picture}
\caption{Experimental scheme of a WW incomplete Mach-Zender interferometer.} %% ALSO FILE fig3.tex SHOULD BE AVAILABLE

\end{figure}
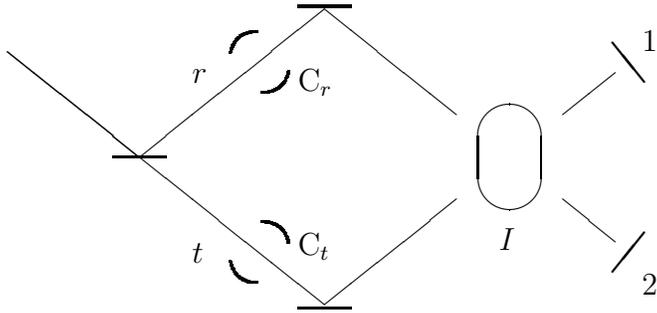

So, in terms of Dirac kets we can schematize the incomplete WW Mach-Zender
interferometer as the following unitary transformations: 
\[
\left| \Psi \right\rangle \stackrel{BS}{\longmapsto }\left| \Psi
\right\rangle _{bs}\stackrel{WW}{\longmapsto }\left| \Psi \right\rangle _{ww}
\]
where the first transformation refers to passage through the beam splitter
(BS) and the second through the which way apparatus (WW). We shall consider 
\begin{eqnarray*}
\left| \Psi \right\rangle _{bs} &=&\left| \psi _r\right\rangle +\left| \psi
_t\right\rangle , \\
\left| \Psi \right\rangle _{ww} &=&\left| \psi _r\right\rangle \left|
r\right\rangle +\left| \psi _t\right\rangle \left| t\right\rangle .
\end{eqnarray*}
The central point which reinforces Bohr's complementarity principle is that 
\[
_{ww}\left\langle \Psi \mid \Psi \right\rangle _{ww}\simeq \left| \psi
_r\right| ^2+\left| \psi _t\right| ^2+2\mathrm{Re}\left\{ \left\langle \psi
_r\mid \psi _t\right\rangle \left\langle r\mid t\right\rangle \right\} 
\]
and so, as we should consider the one bit WW\ states $\left| r\right\rangle $
and $\left| t\right\rangle $ orthogonal to each other, with a WW
interferometer we see no more interference pattern (unless we record
coincidence counts between WW detectors and usual ones).

\section{WW Interferometers in Bohmian Mechanics}

We now treat the example above in Bohmian mechanics, and show nothing
surreal to happen. Our results differ from ESSW ones in reason of the
configuration space we utilize includes the WW one bit information.

As a WW interferometer access also the WW bit, we now should consider the
wave field $\Psi \left( w,\mathbf{x}\right) $, where $w=r,t$ is exactly this
WW bit, and $\mathbf{x}$ has the same meaning as in the simple
interferometer. We can now revisit the three arguments implied to justify
the $r\rightarrow 1$, $t\rightarrow 2$ rule for Bohm trajectories of the
simple interferometer (fig 2):

i) as now we have 
\[
\Psi \left( w,\mathbf{x}\right) =\psi _r\left( r,\mathbf{x}\right) +\psi
_t\left( t,\mathbf{x}\right) , 
\]
there is no more overlap, even when $\mathbf{x}$ describes points in $I$
region, as $w$ assumes different values for each wave packet, there is no
overlap. So, as discussed, quantum potential plays no role and particles
goes like in \emph{free} motion.

ii) If we apply the symmetry operation $\mathbf{R}$ now we just have $\psi
_r\left( r,\mathbf{x}\right) =\pm \psi _t\left( t,\mathbf{Rx}\right) $. The
only conclusion is that there is no \emph{net} flux across the beam splitter
plane, i.e. $\mathbf{v}_{\bot }\left( r,\overline{\mathbf{x}}\right) =-%
\mathbf{v}_{\bot }\left( t,\overline{\mathbf{x}}\right) $, where $\mathbf{v}%
_{\bot }$ denotes the perpendicular to beam splitter component of velocity
vector field, and $\overline{\mathbf{x}}$ denotes invariant points with
respect to $\mathbf{R}$.

iii) As the configuration space has now two slices, trajectories do not
cross, but just passes one ``over'' the other.

In view of this arguments, specially first of them, one can infer (complete
calculations and numerical examples will be given elsewhere) that, \emph{for
WW interferometers}, the correct correspondence is $r\rightarrow 2$, $%
t\rightarrow 1$, exactly as common sense would say is ``the right way''. If
we ignore WW variable, then we must project results and the obtained picture
shows a cross. We should remember it is not a true cross, but a projected
one, which is common result in fluid dynamics (projected flows do not behave
like flows!). In figure 4 we show pictorially how this works.

\begin{figure}[h,t]
\centerline{
\includegraphics [width=2in,height=3in] {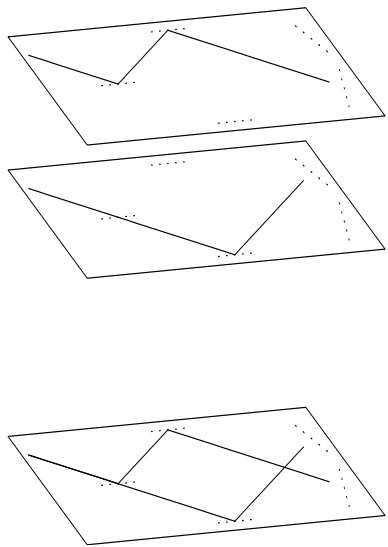}}
\caption{Bohm trajectories in a WW incomplete Mach-Zender interferometer. 
The two upper planes reffers to $w=r,t$ and bottom plane is their 
projection. Configuration space trajectories do not cross, but 
projected trajectories do.}
\end{figure}

It should be stressed that this projection property is related to treat
impure states, or statistical mixtures. For this case, there is no real $%
\Psi $ wave field and Bohmian mechanics does not apply. The author is
presently working in a way of by pass Bell's comment\cite{Bell80} ``%
\textit{in the de Broglie-Bohm theory a fundamental significance is given to
the wave function, and it cannot be transferred to the density matrix}''.
The idea is to diagonalize the density matrix $\rho $ and then consider each
eigenvector independently.

\section{Conclusions}

We have shown Bohmian mechanics to be, not only self consistent, but also to
follow common sense in the example previously considered \textit{%
surrealistic}. We have stressed the main points in the confusion: ESSW had
used Bohm trajectories of a system to analyze a different one. This letter
also made clear how carefully we have to be about configuration space, and
also that, up to now, the only way to avoid confusion, is only ascribe Bohm
trajectories for pure states with a well determined wave field.

\begin{ack}

I would like to thank Professor M. O. Scully for introduce the theme and
providing some references. The author also thanks the organizers of Escuela
Latinoamericana de F\'{\i}sica, held in Mexico, and where part of this work
was done, for hospitality. Finally, I acknowledge D.A. Manoel and
M. Ferreira for the eforts in providing necesary references, and Doctor 
T. O. Carvalho for assistence with pictures.

\end{ack}

\end{document}